\documentstyle[twoside,fleqn,npb,epsfig]{article}
\newcommand{\AmS}{{\protect\the\textfont2
  A\kern-.1667em\lower.5ex\hbox{M}\kern-.125emS}}

\title{Loops and legs beyond perturbation theory\thanks{Contributed to the {\em Loops and Legs in Quantum Field Theory} conference, April 9--14, 2000,
 Bastei-K\"onigstein, Germany.}}

\author{Adrian Ghinculov\address{Department of Physics and Astronomy, UCLA,
                                 Los Angeles, California 90095-1547, USA}%
        \thanks{Work supported by the US Department of Energy}
        and
        Thomas Binoth\address{Laboratoire d'Annecy-Le-Vieux de Physique 
         Th\'eorique LAPTH,\\
         Chemin de Bellevue, B.P. 110, F-74941, Annecy-le-Vieux, France}%
         \thanks{UMR 5108 du CNRS, associ\'ee \`a l'Universit\'e de Savoie.}}

\begin{document}
\begin{abstract}
%\begin{flushright}
%UCLA????\\
%LAPTH-797-00\\
%May 2000
%\end{flushright}
%\vspace{2cm}
Within the non-perturbative $1/N$ expansion, we discuss numerical 
methods for calculating multi-loop Feynman graph needed to derive physical
scattering amplitudes. 
We apply higher order $1/N$ methods to the scalar sector of the standard model,
%discuss the treatment of 
%the tachyon pole, 
and show the existence of a mass saturation effect.
The mass saturation has direct implications for future searches at 
the LHC and at possible muon colliders.
\end{abstract}

\maketitle

\section{INTRODUCTION}

There has been a lot of work lately in extending 
the calculation of radiative corrections to higher loop-orders and to 
multi-leg processes. The proceedings of this 
Workshop are suggestive for the complexity this field has reached
since the pioneering work of Martinus Veltman and Gerardus 't Hooft.
Higher order QCD radiative corrections are necessary for 
interpreting the experimental data which the LHC experiments will provide.
Higher order electroweak  corrections are needed for interpreting
precision electroweak measurements. 
Due to the time frame for the construction of future colliders,
precision measurements will continue 
to be a main tool for new physics searches.

In this contribution we would like to discuss the use of multi-loop Feynman
graphs for recovering results which are normally not within the scope 
of perturbation theory, namely deriving scattering amplitudes in the
case of strongly coupled field theories.

A central question in particle physics is how is the electroweak symmetry
broken. The experimental data currently available is compatible with a
standard model scalar sector, and favors a light Higgs boson. However, 
additional degrees of freedom beyond the standard model have the 
potential to shift these fits. It is not at all excluded that the 
LHC experiments will see a scalar resonance
at a considerably higher energy.

At strong coupling, the 
scalar sector was insufficiently explored. Qualitatively,
one may expect new phenomena to appear: a Higgs particle strongly coupled to
the vector bosons and to itself, anomalous vector boson self-interactions, and
possibly the appearance of a spectrum of additional resonances in the 
scalar sector. Perturbative calculations loose their predictive power
when radiative corrections blow up in higher loop orders, and
the dependency on the renormalization scheme becomes substantial \cite{scheme}.

Ideally, one needs a solution for the scalar sector which is valid
at strong coupling as well as at weak, and which is free of 
renormalization scheme uncertainty. This can be accomplished by explicitly 
summing up all loop orders within a non-perturbative $1/N$ expansion.
If a solution of sufficient accuracy is desired, such that it can be used
in phenomenological calculations to compete with ordinary perturbation theory,
then the $1/N$ expansion must be carried out at higher order.

\section{THE $1/N$ SOLUTION}

We are interested in a solution of the scalar sector of the standard model
when the self-coupling of the Higgs field becomes strong. The non-perturbative
effects are thus given by the scalar field self-interaction, while the gauge
coupling remains perturbative. It is then natural to resort to the 
equivalence theorem to relate amplitudes involving longitudinal electroweak 
gauge bosons to the corresponding amplitudes involving would-be Goldstone 
bosons. 
The problem is being reduced to solving a linear sigma model 
non-perturbatively. 

\subsection{The auxiliary field formalism}

It is sufficient
to consider the scalar sector alone, which is an $SU(2)$-symmetric 
linear sigma model. In the following, the scalar sector is extended
to an $O(N)$-symmetric sigma model --- with the standard model case  
recovered for $N=4$ --- such that an expansion in powers of
$1/N$ can be performed:

\begin{eqnarray}
  {\cal L}_1 & = &   \frac{1}{2}           
               \partial_{\nu}\Phi_0 \partial^{\nu}\Phi_0 
             - \frac{\mu_0^2}{2}      \Phi_0^2 
	     - \frac{\lambda_0}{4! N} \Phi_0^4 
 ~~ , \nonumber \\
 \Phi_0 & \equiv & \left( \phi_0^1, \phi_0^2, \dots , \phi_0^N \right)
\end{eqnarray}

From the Lagrangian above, it is easy to derive scattering amplitudes at
leading order in $1/N$. Typically, this involves summing up a geometric 
series of one-loop bubble diagrams which have Goldstone bosons in the loop.

However, the auxiliary field formalism which was proposed
in ref. \cite{coleman} proves to be most useful in organizing sub-leading 
orders in the $1/N$ expansion in a diagrammatically manageable way.
It consists in introducing an additional unphysical field $\chi$ in the
Lagrangian: 

\begin{eqnarray}
  {\cal L}_2 & = & {\cal L}_1 + \frac {3 N}{2 \lambda_0} 
                 (\chi_0 - \frac{\lambda_0}{6 N} \Phi_0^2 - \mu_0^2)^2 
	\nonumber \\	
           & = & 
    \frac{1}{2} \partial_{\nu}\Phi_0 \partial^{\nu}\Phi_0 
  - \frac{1}{2} \chi_0 \Phi_0^2 
  + \frac{3 N}{2 \lambda_0} \chi_0^2
	\nonumber \\	
           &  &	   
  - \frac{3 \mu_0^2 N}{\lambda_0} \chi_0 + const. 
\end{eqnarray}

Because the equation of motion for the auxiliary field $\chi$ is 
a constant, this addition does not change the dynamics.
Green's functions having Higgs and Goldstone bosons on the legs 
are the same, whether calculated with the Feynman rules
given by ${\cal L}_1$ or by ${\cal L}_2$.

The Feynman rules, on the other hand, are changed. ${\cal L}_1$
contains the trilinear and quartic vertices $\sigma\pi\pi$,
$\sigma\sigma\sigma$, $\sigma\sigma\sigma\sigma$, 
$\sigma\sigma\pi\pi$, and $\pi\pi\pi\pi$. Here $\sigma$ and $\pi$
are the massive and massless modes stemming from Lagrangian ${\cal L}_1$
after spontaneous symmetry breaking, respectively. ${\cal L}_2$ contains
only the trilinear couplings $\chi\sigma\sigma$ and $\chi\pi\pi$. 

This simplifies enormously the topological classification of  
Feynman diagrams according to their power of $1/N$ beyond leading order. 
The reader can easily convince himself of the utility of the auxiliary 
field formalism by writing down the diagrams which contribute to 
Goldstone-Goldstone scattering at NLO in $1/N$ in both formalisms.

\subsection{Tachyonic regularization}

By summing up the chains of one-loop bubble self-energy insertions,
one encounters an ultraviolet renormalon. This gives rise to an additional,
tachyonic pole in the propagators, apart from the expected
physical spectrum containing one Higgs boson and $N-1$ Goldstone modes.

By direct evaluation of the leading order contribution in 
$1/N$ to the two-point functions, one obtains the following propagators
\cite{coleman,other}:

\begin{eqnarray}
  D_{\sigma \sigma}(s) &=&                        \frac{i         }{s - m^2(s)} \nonumber \\
  D_{\chi \chi}(s)     &=& \frac{1}{      N  v^2} \frac{i s m^2(s)}{s - m^2(s)} \nonumber \\
  D_{\chi \sigma}(s)   &=& \frac{1}{\sqrt{N} v  } \frac{i   m^2(s)}{s - m^2(s)} \nonumber \\
  D_{\pi_i \pi_j}(s)   &=& \frac{i}{s}  \delta_{ij}  ~~~ ,
\end{eqnarray}
where

\begin{eqnarray}
  m^2(s) & = & \frac {v^2}{\frac{3}{\lambda} + \hat{\alpha}^{(0)}(s)} 
  \nonumber \\
         & \equiv &
            \frac {v^2}{\frac{3}{\lambda} - \frac{1}{32 \pi^2} 
                        \log{\left(-\frac{s+i \eta}{\mu^2}\right)}} ~~~ .
\end{eqnarray}     
Here $\hat{\alpha}^{(0)}(s)$ 
is the ultraviolet finite part of the one-loop self-energy bubble diagram,
with a Goldstone boson in the loop. $\mu$ is the ultraviolet subtraction scale.

The propagators contain an Euclidian pole in the ultraviolet region
at an energy $s=-\Lambda_t^2$ given by the following transcendental equation:

\begin{equation}
    \frac{v^2}{\Lambda_t^2} 
  - \frac{1}{32 \pi^2} \log{\left( \frac{\Lambda_t^2}{\mu^2} \right)}
  + \frac{3}{\lambda} = 0   ~~~~~ .
\end{equation}
The tachyon scale is in the ultraviolet region for low values of the
coupling, and tends to move towards low energy
when the coupling is increased.

In higher order $1/N$ corrections,
the Euclidian pole appears in the loop momentum integration.
This induces causality violating contributions, even though these 
effects are numerically small as long as the tachyon scale is high enough.
For this reason, it is necessary to investigate the origin of this singularity
and find a way for dealing with it.

At any finite order of perturbation theory, the tachyon pole does 
not exist. It is an artifact of the bubble diagram summation. 
In the process of summing up all loop orders,
an ambiguity is present, however. The summation of the perturbative series
determines the result only up to functions which vanish in perturbation 
theory, such as $e^{-1/\lambda}$. Because the residuum of the tachyonic pole
is precisely such a function which vanishes in perturbation theory, its 
presence or absence is completely arbitrary and cannot be determined within
perturbation theory. For this reason, it is justified to restore causality
by minimally subtracting the tachyon at its pole since the original information
stemming from Feynman diagrams remains unchanged \cite{1ovn}:
\begin{eqnarray}
  D_{\sigma \sigma}(s) &=& i \left[   \frac{1     }{s - m^2(s)}
                                 - \frac{\kappa}{s + \Lambda_t^2} \right] \nonumber \\
  D_{\chi \chi}(s)     &=& \frac{i s}{N  v^2} 
                          \left[   \frac{m^2(s)     }{s - m^2(s)}
                                 + \frac{\kappa \Lambda_t^2}{s + \Lambda_t^2} \right] \nonumber \\
  D_{\chi \sigma}(s)   &=& \frac{i}{\sqrt{N} v}  
                          \left[   \frac{m^2(s)     }{s - m^2(s)}
                                 + \frac{\kappa \Lambda_t^2}{s + \Lambda_t^2} \right]
 ~~,
\end{eqnarray}
were 
 $ \kappa = [ 1 + \Lambda_t^2/(32 \pi^2 v^2) ]^{-1} $
%\begin{equation}
%  \kappa =  \frac{1}{ 1 + \frac{\Lambda_t^2}{32 \pi^2 v^2} }
%\end{equation}
is the residuum of the tachyonic pole. 

The tachyonic regularization can be seen as a prescription for summing up
the  perturbative series in a way which preserves causality.

\subsection{$1/N$ renormalization}

Performing renormalization at NLO in the $1/N$ expansion involves the 
treatment of ultraviolet divergences of diagrams with various numbers of loops.
One way of keeping track of various loop order counterterms is to group
them into $1/N$ counterterms. Each of the $1/N$ order counterterms 
$\delta \lambda$, $\delta t$, $\delta t_{\chi}$, 
$\delta Z_{\pi,\sigma,\chi}$ is a power series in the coupling 
constant $\lambda$ \cite{1ovn}:
\begin{eqnarray}
  \frac{3}{\lambda_0}       &=&   \frac{3}{\lambda} + \Delta \lambda   
\nonumber \\
  & \equiv &\frac{3}{\lambda} + \delta \lambda^{(0)} 
                             + \frac{1}{N} \delta \lambda
                             + {\cal O}\left(\frac{1}{N^2}\right)
\nonumber \\
  \frac{3 \mu_0}{\lambda_0} &=& - \frac{v^2}{2} (1 + \Delta t)
\nonumber \\
  & \equiv & - \frac{v^2}{2} \left[ 1 + \frac{1}{N}\delta t 
                               + {\cal O}\left(\frac{1}{N^2}\right) \right]
\nonumber \\
  \phi^i_0 &=&  \pi_i  Z_{\pi}  ~~~~~~~~~~~~~~ , ~~  i=1, \dots ,N-1         
\nonumber \\
  & \equiv & \pi_i \left[ 1 + \frac{1}{N} \delta Z_{\pi}
                          + {\cal O}\left(\frac{1}{N^2}\right)  \right] 
\nonumber \\
  \phi^N_0 &=&  \sigma Z_{\sigma} + \sqrt{N} v                         
\nonumber \\
  & \equiv & \sigma \left[ 1 + \frac{1}{N} \delta Z_{\sigma}
                          + {\cal O}\left(\frac{1}{N^2}\right)  \right] 
         + \sqrt{N} v
\nonumber \\
  \chi_0   &=&  \chi   Z_{\chi} + \hat \chi +\Delta t_{\chi}
\nonumber \\
  & \equiv & \chi \left( 1 + \frac{1}{N} \delta Z_{\chi}  \right) 
          + \frac{v^2}{N} \delta t_{\chi} + {\cal O}\left(\frac{1}{N^2}\right)
\end{eqnarray}

\begin{figure}
%  \begin{center}
  \epsfxsize = 5.7cm \epsffile{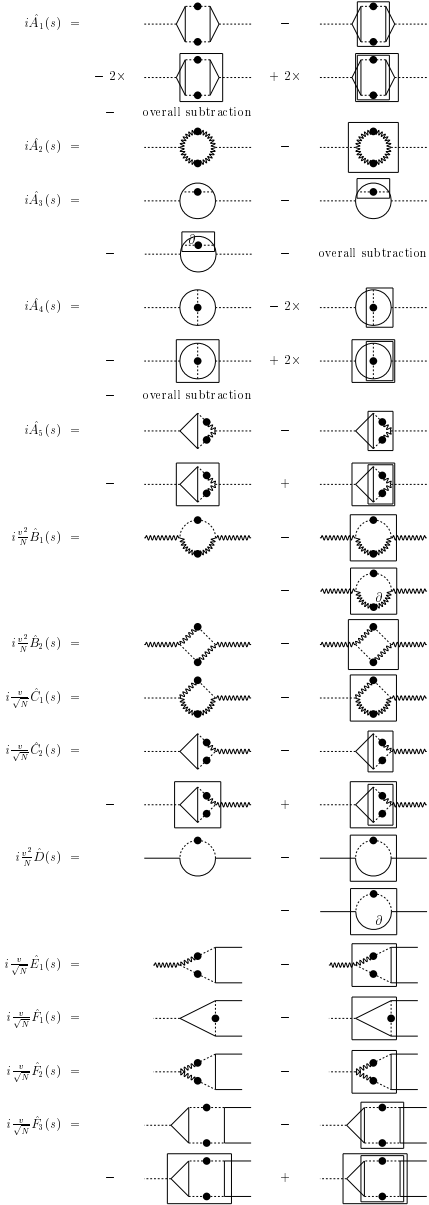}
%  \end{center}
\caption{{\em Ultraviolet subtractions
              of all NLO $1/N$ two- and three-point diagrams of the sigma
              model. The blob on the propagators denotes the resummed chain 
              of one-loop Goldstone bubble diagrams. The solid line denotes
              Goldstone fields, the wavy line is the Higgs field, and the
              dashed line is the auxiliary field.}}
\end{figure}

In principle, these counterterms are sufficient to absorb the
UV divergences of all diagrams of NLO in $1/N$.

When calculating  $1/N$ diagrams at NLO, we  resort to
numerical integration. The ultraviolet divergence of the diagram must
be removed before the numerical integration can be carried out, but the
explicit $\epsilon$ expansion and isolation of UV poles is  cumbersome
because of the complexity of the diagrams. For this reason we perform
a BPHZ-type renormalization \cite{1ovn}. 
We subtract the divergences of the diagrams
according to the forest formula, as shown in figure 1. The subtracted 
expressions, being finite in the ultraviolet, can be integrated numerically.
For a given physical process, the renormalization program then means to
combine all UV subtraction terms from various diagrams with each other.
Most subtractions cancel with each other, 
and the small remaining set of subtractions
is trivially absorbed into the $1/N$ local counterterms above. We have 
checked that all UV subtractions are polynomial, as they ought to be in 
order to be absorbed into local counterterms.

\subsection{Numerical solution}

The finite expressions depicted in figure 1 are calculated
by numerical integration because so far no analytical approach 
exists for dealing with this type of diagrams. 

For a numerical solution, it is advantageous to identify first all
loop integrations which can be performed analytically. These are  
associated with closed Goldstone loops. For the diagrams shown in figure 1,
it is possible to perform analytically all integrations except for one final
loop integration. The final integration involves the resummed and 
tachyonically subtracted propagators of eqs. 6 and form factors from
one-loop triangle or box diagrams involving massless Goldstone bosons 
in the loop. The final loop integration needs to be performed numerically.
It can be reduced to a two-fold integral with the methods of refs. 
\cite{3loop,1ovn}, which do not resort to Feynman parameters.

\subsection{Scheme independence of physical predictions}

For a given order in the expansion parameter $1/N$, the Feynman diagrams
of all loop orders are explicitly summed up. For this reason, the final 
result for a physical scattering amplitude is free of 
renormalization scheme
dependence. In usual perturbative calculations, a renormalization scheme
ambiguity is present because of the truncation of the perturbative 
expansion; this ambiguity is of higher order in the coupling constant.

We note that subtracted diagrams of the type 
shown in figure 1 are not
individually free of renormalization scheme dependency. The internal 
subtractions which are performed for making the integrals
finite in the ultraviolet, are calculated at a given 
subtraction point. This defines an intermediary renormalization scheme. 

This renormalization scheme dependence cancels out only in physical results
such as scattering amplitudes. In the following section we give the results
for two scattering processes. We have checked explicitly that the two functions
involved, $f_1$ and $f_2$ of eqns. 9, 
are independent of the subtraction point.

\section{$\mu^+\mu^-$ COLLIDERS AND THE HIGGS MASS SATURATION EFFECT}

Recently, feasibility studies for muon colliders attracted quite
some attention. A muon collider would be an ideal $s$-channel
Higgs factory. For a heavy Higgs boson, two processes dominate:
$\mu^+\mu^- \rightarrow H \rightarrow t\bar t$ and
$\mu^+\mu^- \rightarrow H \rightarrow W^+_LW^-_L, Z_LZ_L$.

Within the $1/N$ expansion, the amplitudes for these processes at
NLO are given by the following expressions \cite{muoncoll}:
\begin{eqnarray}
{\cal M}_{f\bar{f}} & = &
\frac{1}{s - m^2(s) \left[ 1 - \frac{1}{N} f_1(s)  \right] }
  \nonumber  \\
{\cal M}_{ZZ} & = &
\frac{m^2(s)}{\sqrt{N} v}
\frac{1 - \frac{1}{N} f_2(s) }{s - m^2(s) \left[ 1 - \frac{1}{N} f_1(s)  \right] }
~~.
\end{eqnarray}
Here, the correction functions $f_1$ and $f_2$ are given by a combination
of the two- and three-point functions defined in figure 1 (see \cite{1ovn}):
\begin{eqnarray}
f_1(s)  & = &
           \frac{m^2(s)}{v^2} \, \hat{\alpha}(s)
       + 2 \hat{\gamma}(s)
       +   \frac{v^2}{m^2(s)} \left[ \hat{\beta}(s) \right.
  \nonumber  \\
    & &   \left.  - 2 \frac{s-m^2(s)}{v^2} \left( \delta Z_{\sigma} - \delta Z_{\pi} \right)
                              \right]
  \nonumber  \\
f_2(s)  & = &
           \frac{m^2(s)}{v^2} \, \hat{\alpha}(s) 
       +   \hat{\gamma}(s)
  \nonumber  \\
    & &       -   \hat{\phi}(s)
       -   \frac{v^2}{m^2(s)} \hat{\eta}(s)  ~~.
\end{eqnarray}

\begin{figure}[t]
  \epsfxsize = 6cm 
%  \begin{center}
    \epsffile{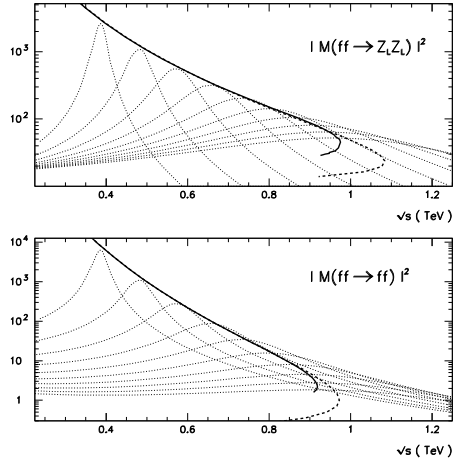}
%  \end{center}
\caption{{\em The Higgs line shape at a $\mu^+\mu^-$ Higgs factory.
              We marked the position of the maxima of the resonances
              (solid line for the $1/N$ result and dashed line for the
               perturbative result at two-loop).}}
\end{figure}

In figure 2 we give numerical results for these expressions. We
plot the shape of the Higgs resonance for various strengths of the
coupling. From the positions of the resonance maxima, it can be seen
clearly that when the coupling is increased, a mass saturation effect
appears. The position of the resonance's peak does not increase above 
a saturation value just under 1 TeV. The precise resonance shape and 
implicitly the saturation value are process dependent. 

For comparison,
we indicate in figure 2 the corresponding peak maxima  which would be obtained 
by using usual perturbation theory at NNLO. A saturation is present
qualitatively in this curve, too. However, it should be noticed that
the perturbative curve is affected by large radiative corrections and large
scheme uncertainties at such large coupling, and therefore is not reliable
in the saturation region.

\section{SATURATION EFFECT AT THE LHC}

The main Higgs production mechanism at the LHC is gluon fusion
which proceeds through a top loop \cite{glover}. The existing perturbative
analyses indicate that the vector boson fusion becomes competitive 
at an energy of the order of 1 TeV.

In the presence of the saturation effect, the observation
of a heavy Higgs resonance at the LHC will be different. In the
saturation zone, the mass of the resonance remains more or less the same
while the width increases with the coupling. The resonance becomes flatter
and more difficult to detect.

\begin{figure}[t]
  \epsfxsize = 6cm
%  \begin{center}
    \epsffile{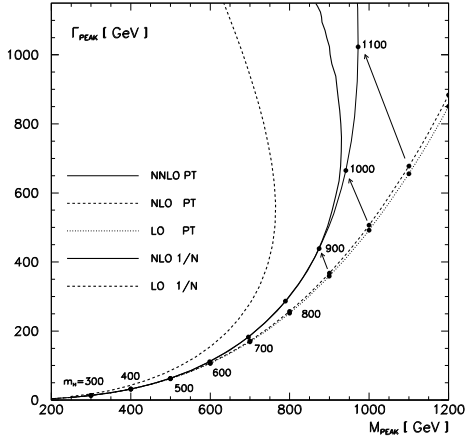}
%  \end{center}
\caption{{\em The current knowledge of the Higgs width at strong coupling
from perturbation theory an from the $1/N$ expansion.
The parameters $M_{PEAK}$
and $\Gamma_{PEAK}$ are extracted from the position and the height 
of the Higgs resonance in fermion scattering as if the resonance was 
of Breit-Wigner type. For
the perturbation theory curves we give the corresponding values
of the on-shell mass parameter $m_H$.}}
\end{figure}

In figure 3 we show a plot of the Higgs
width as a function of the Higgs mass. The width and mass used in these
plots, $M_{PEAK}$ and $\Gamma_{PEAK}$, are defined from the line shape
of the resonance as seen in fermion-fermion scattering. They are derived 
from the position and height of the resonance as if it were of Breit-Wigner
type. In this picture we show the calculations available so far in 
usual perturbation theory (LO, NLO, NNLO) \cite{2loop}, 
and in the $1/N$ expansion (LO and NLO) \cite{1ovn}. 
It can be seen that the two expansions converge nicely
towards each other and display a mass saturation.

We studied the discovery potential of the LHC in the presence 
of the saturation effect. In ref. \cite{lhc} we performed a Monte-Carlo
simulation for the LHC. Only leptonic channels were considered, namely
the ``golden plated'' channel $(l^+l^-)(l^+l^-)$, and the 
$l^+l^-\nu\bar \nu$ channel. When neutrinos in the final state are involved,
the Higgs resonance appears as a Jacobian peak in the distribution of missing
$p_T$. We included the gluon fusion process together with the relevant 
background. The strong interacting Higgs correction was included by using
the NNLO perturbative calculation - it results into a faster Monte-Carlo
while being a fair approximation of the non-perturbative $1/N$ result.

We used usual assumptions about the LHC energy and luminosity, and asked
for a $5 \sigma$ effect with respect to the background. The discovery 
potential estimated for the ``golden plated'' channel corresponds then 
to an on-shell Higgs mass of 830 GeV. The missing $p_T$ channel reaches
up to an on-shell Higgs mass of 1030 GeV. Note that these values for the
on-shell mass deviate considerably from the actual position of the resonance.
The actual peak can be read from figure 3, where the on-shell mass
is mapped onto the saturation curve.

\section{CONCLUSIONS}

The combinatorial structure of the sigma model makes possible
an explicit calculation of the non-perturbative $1/N$ expansion
at NLO. 

The $1/N$ solution is valid at strong coupling as well as at weak
coupling. It is also independent of the intermediate renormalization
scheme which is being used. At NLO it also provides for ultraviolet
finite wave function renormalization constants.

The non-perturbative solution of the sigma model is relevant for the
standard model via the equivalence theorem. It implies that a mass
saturation effect is present in the scalar sector. When the Higgs
self-interaction is increased, the mass of the resonance increases
only up to a maximum value under 1 TeV, while the width increases
continuously. 

An interesting question is how cutoff effects stemming from 
new physics at higher energies can modify
the $1/N$ solution. This clearly deserves further investigation.

\vspace{.2cm}
{\bf Acknowledgement}
The authors are grateful to the Organizers of the 
{\em Loops and Legs in Quantum Field Theory 2000} conference for
having organized an exciting and stimulating meeting.
The work of T.~B. 
is supported by the EU Fourth Training Programme  
''Training and Mobility of Researchers'', Network ''Quantum Chromodynamics
and the Deep Structure of Elementary Particles'',
contract FMRX--CT98--0194 (DG 12 - MIHT).

\end{document}